# Extreme Low-Frequency Ultrathin Acoustic Absorbing Metasurface

Krupali Donda[1], Yifan Zhu[1], Shi-Wang Fan[1], Liyun Cao[1], Yong Li[2] and Badreddine Assouar[1]*

[1]*Institut Jean Lamour, CNRS, Université de Lorraine, Nancy, France*

[2]*Institute of Acoustics, School of Physics Science and Engineering, Tongji University,*

*Shanghai 200092, China*

**Abstract:**

We introduce a multi-coiled acoustic metasurface providing a quasi-perfect absorption (reaching 99.99% in experiments) at extremely low-frequency of 50 Hz, and simultaneously featuring an ultrathin thickness down to $\lambda/527$ (1.3 cm). In contrast to the state of the art, this original conceived multi-coiled metasurface offers additional degrees of freedom capable to tune the acoustic impedance effectively without increasing the total thickness. We provide analytical derivation, numerical simulation and experimental demonstrations for this unique absorber concept, and discuss its physical mechanism which breaks the quarter-wavelength resonator theory. Furthermore, based on the same conceptual approach, we propose a broadband low-frequency metasurface absorber by coupling unit cells exhibiting different properties.

**Keywords:** Acoustic metasurfaces/metamaterials; Low-frequency absorption; Ultra-subwavelength thickness; Coiling-up space geometry

**Corresponding author:**

*badreddine.assouar@univ-lorraine.fr

Due to the weak intrinsic dissipation of conventional materials in the low-frequency region, the perfect absorption of sound at low frequency (<100Hz) is still a scientific challenge [1]. To enhance the dissipation, it is necessary to increase the energy density inside the relevant material by some means like through resonances [2]. In the past two decades, the explosion of interest in developing artificial resonant structures like acoustic metamaterials [3-6] and metasurfaces [7-12] was drastically increased owing to their exotic capabilities of manipulating sound waves and their deep subwavelength thickness [13, 14]. The remarkable characteristics of acoustic metamaterials such as negative mass density, refractive index, double negativity, and controlled anisotropy have attracted massive research in the field of developing deep subwavelength sized acoustic devices for low-frequency applications such as negative refraction [13], deep subwavelength focusing/imaging [14], cloaking[15], perfect absorbers with ultrathin thickness [12,14-22] etc. In related researches, perfect metasurface absorbers have received considerable attention, and variety of acoustic metasurfaces designs have been designed for the applications in noise control at low-frequency regime, which could be used in aircraft, locomotives, automobiles, machines, and buildings [13, 17-24].

Ultra-thin acoustic absorber metasurface is generally based on a single or hybrid resonant system [5-12; 16-23]. The hybrid resonant system [25] usually has a broader bandwidth than a single resonant one. One interesting way to design the perfect acoustic absorber is to use an ultrathin decorated membrane with thin air chamber. Such a system was reported by Ma et al. [7] to form hybrid resonance capable of to realizing complete absorption at low-frequency range [8, 9] with a thickness of about ~1/133 of the operating wavelength. However, this design requires controlled tension of the membrane which may cause fabrication challenges and durability issues. Jimenez et al. [17] reported an ultra-thin omnidirectional perfect acoustic absorber (with the thickness of ~1/88 of the operating wavelength) using the mechanism of slowing sound and critical coupling. For the broadband performance, Yang et al. [25] proposed optimal sound-absorbing structures based on multiple resonant tubes to achieve highly efficient absorption within the frequency range of 350Hz-3000Hz, using a structure thickness of $\lambda/10$ for the lowest working frequency. Among all reported structures, one promising concept is to introduce a medium with a high refractive index. It can be realized by using the coiled-up space geometry [13, 26]. Such coiled structures provide a more direct way to drastically shrink the thickness of the system. A prototype of such metasurface was proposed by Li. et al. [12] with the structural thickness being $\lambda/223$, showing the ultra-thin property of coiled acoustic metasurface. The disadvantage of this system is that the absorption depends on the length of the coiled channel and strictly requires quarter wavelength channel length for the full absorption, creating difficulties in adjusting the absorption frequency and bandwidth. To improve this design, Huang et al. [24] reported a spiral metasurface (thickness being $\lambda/100$) with adding an aperture with the coiled structure to have an additional degree of freedom to tune the absorption frequency.

To overcome these limitations in terms of both absorption frequency and associated metasurface thickness, in this letter, we introduce an original absorbing physical mechanism enabling the simultaneous achievement of extreme low-frequency absorption with ultra-subwavelength metasurface thickness. Unlike the conventional mechanism based on quarter wavelength resonator by which we can shrink a meta-structure to make it thin while conserving the same operating frequency, our reported mechanism breaks this physics limitation by introducing a novel concept based on a multi-coiled metasurface (MCM). This achieved concept is composed of perforated plate, coplanar coiled chamber with an aperture and labyrinthine passages. The multi-coiled geometry is formed by adding the labyrinthine passages inside the coiled chamber. The advantages of this concept are first, to fully utilize the space inside the coiled chamber, and second to have an additional degree of freedom which helps to shift the resonance towards extreme low-frequency without increasing the total thickness of the metasurface. It also weakens the dependence of the resonance frequency on the total channel length. As we will demonstrate, absorbing metasurface operating at 50 Hz with a total thickness down to an unprecedented $\lambda/527$ is achieved. We first will provide an analytical model based on the equivalent circuit method [27, 28] to derive the absorption coefficient. Then, a numerical simulation based on finite element model will be implemented to be compared and discussed against the theoretical approach. Finally, we will conduct experimental fabrication, measurement and analysis of the conceived metasurfaces to provide a proof-of-concept of our finding. We, actually, will present the first demonstration of an extreme low-frequency perfect absorber operating at 50Hz and having only 1.3cm thickness. Then, we will provide a second demonstration of a broadband absorber of the same thickness and exhibiting a high absorption over 85%.

In general, acoustic absorption can be defined as dissipation of sound energy by the absorbing system or material. The acoustic absorption coefficient, $\alpha$, of such a system can be determined by its normal acoustic impedance, $Z$ and this relationship can be given by [28],

$$\alpha = 1 - \left|\frac{Z - \rho_0 c_0}{Z + \rho_0 c_0}\right|^2 \tag{1}$$

where, $\rho_0$ is the static air density and $c_0$ is the sound speed. Equation (1) indicates that perfect absorption requires the impedance matching between the air and the metasurface. i.e., $\text{Re}(Z) = \rho_0 c_0$ and $\text{Im}(Z) = 0$.

Figure 1(a) shows the structural assembly of the coplanar chamber, embedded aperture, and labyrinthine passages while Fig. 1(b) illustrates the detailed view of the MCM. The circular aperture is fixed with the perforated plate at the opening of the coiled channel. It is observed that the longer aperture contributes to shift the resonant frequency in the lower region without increasing the length of the coil. A total of 11 labyrinthine passages of different size are embedded into the coplanar chamber which essentially force incoming waves to propagate through passages that are much longer than their external dimension.

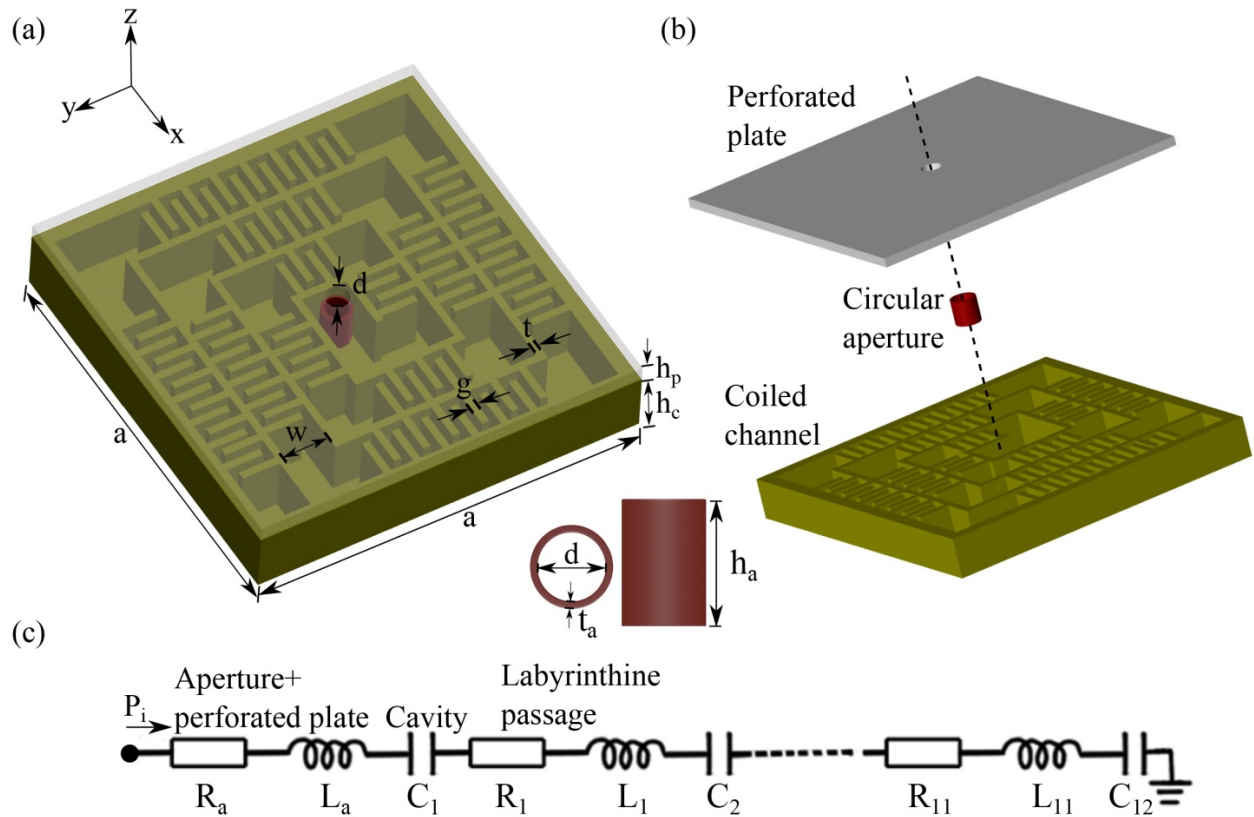

**Figure 1**: Schematic of a multi-coiled metasurface absorber (cross area $a \times a$) composed of a coiled channel, labyrinthine passages, aperture and perforated plate (a) the structural assembly of coplanar channel ($w$ is the width of channel, $h_c$ is the height of the coiled chamber, $t$ is the thickness of the wall), embedded aperture (height $h_a$, thickness of wall $t_a$ and diameter $d$), labyrinthine passages (width of channel $g$), and perforated plate (hole diameter $d$, thickness $h_p$). The periodicity in both $x$-$y$ directions is $a$. (c) the effective circuit model of the unit cell. The normal incident wave propagates along the z direction and penetrates into the coiling chamber through the aperture. (b) detailed view of the structure

The MCM can be theoretically analyzed using a lumped RLC circuit model. The equivalent electrical circuit of the system is shown in Fig. 1(c). The circular aperture, perforated plate, and labyrinthine structures contribute essentially to the acoustic resistance and inductance of the structure, while the cavities (gaps between two labyrinthine passages) contribute essentially in acoustic capacitance. Herein, we will give the analytical derivations of total acoustic impedance of the system. Acoustic impedance per unit length of an infinite aperture can be calculated from the Crandall's theory [29]. In our case, the aperture is relatively long with height, $h_a = $ 9mm and then thermal-viscous loss needs to be considered. Considering the effect of thermal-viscous loss, the acoustic resistance $R_a$ and inductance $M_a$ provided by the long circular aperture contributing to the MCM can be given by [24],

$$R_a = \frac{4A}{\pi d^2}\left[\sqrt{2\omega\rho_0\eta} - \frac{2\rho_0 c \sin\left(\frac{k_c h_a}{2}\right)}{\omega(\gamma - (\gamma-1)\Psi_h)\Psi_v}\right] \quad (2)$$

$$M_a = \frac{4A\rho_0\delta_i}{\pi d^2} \quad (3)$$

where, $\omega$ is the angular frequency, $\eta$ is dynamic viscosity of the air, $k_c$ is the complex wave number, $h_a$ is the height of aperture, $\Psi_h$ and $\Psi_v$ are the functions of thermal and viscous fields in the aperture [29], respectively. $A = a^2$ is the area of the integral metasurface and $\delta_i$ is the end correction. In our case, top part of the aperture is attached with the perforated plate and embedded inside the coiled channel. Then, the end correction $\delta_i$ can be given by, $\delta_i = [1 + (1 - 1.25\epsilon)] \times \delta_0/2$, where $\epsilon = d/\min(w, h_c)$ and $\delta_0 = 0.85d$.

The introduction of labyrinthine structures divides the whole channel geometry into several narrow coiled passages and cavities. The width of labyrinthine passages is *g*. The acoustic resistance and inductance for these passages are calculated by [27]:

$$R = \frac{h_c}{\pi d^3}\sqrt{2\mu\omega\rho_0} + 2\frac{\sqrt{2\mu\omega\rho_0}}{\pi d^2} + \frac{\rho_0 c_0}{\pi d^2}\left[1 - \frac{2J_1(2kd)}{2kd}\right] \quad (4)$$

$$M = \frac{l_b\rho_0}{S} \quad (5)$$

where $l_b$ is the total length and *S* is the total area of the channel, the air density is $\rho_0$=1.21kg/m³, $J_1$ is the Bessel function of the first kind, the sound speed is $c_0$=343m/s and $\omega=2\pi f$ is the angular frequency. The volumes of all 12 cavities are different and their acoustic capacitance is given as [28],

$$C = \frac{V}{\rho_0 c_0^2} \quad (6)$$

where *V* is the volume of the cavity. Since the resistor, inductor, and capacitor are connected in series, the total acoustic impedance, *Z* of the MCM can be expressed by,

$$Z = R_a + \sum_{i=1}^{11} R_i + j\left(\omega M_a + \sum_{i=1}^{11}\omega M_i - \sum_{i=1}^{12}\frac{1}{\omega C_i}\right) \quad (7)$$

The incident acoustic wave ($\lambda \gg a$) along the *z*-direction enters into the coiled chamber via aperture as shown in Fig. 1. Due to the energy dissipation, the sound energy can be highly absorbed at the resonant frequency. To obtain the absorption spectra, we perform numerical calculations with the Acoustic-Thermoviscous Acoustic Interaction module (frequency domain) of COMSOL Multiphysics v5.4 [30]. The effect of the viscous friction and the heat transfer is concluded in the linearized compressible Navier-Stokes equation, the continuity equation, and the energy. Hard boundaries are imposed on the interfaces between air and solid due to the large impedance mismatch between air and solid materials [31]. The absorption coefficient, α is expressed as $α = 1-|r|^2$ with the reflection coefficient $r = (Z-1)/(Z+1)$. The perfect absorption is achieved at 50Hz with the total thickness of the system, $h_c+h_p$=13mm which is 1/527 of the working wavelength. To our best of knowledge, this is by far, the thinnest acoustic metasurface achieved until now for a low frequency acoustic absorption at 50Hz. In order to understand the absorption physical mechanism, acoustic pressure distribution inside the metasurface is shown at the resonant frequency (50Hz) in Fig. 2 (c). The acoustic pressure is higher in deeper regions of the multi-coiled structure at the resonance case, whose distribution is similar to previous designs with gradient channel [32-33]. Acoustic pressure amplitude difference between different coiled passages drives the air-flows in the coiled passage [33], making the whole channel strongly rubbed by the acoustic wave, so that the incident acoustic energy is rapidly attenuated inside.

Regarding the experiments, we fabricate the absorbing metasurface from polylactic acid (PLA) material by 3D printing technology. The sound speed and the density of the used PLA are $c$=1200m/s and $ρ$=2700kg/m$^3$, respectively. Photographs of the fabricated MCM are shown in Fig.2(a). Figure 2(b) shows the experiment setup for the absorption measurement. A lab-made impedance tube (inner size is of 10cm × 10cm), two Brüel & Kjær 1/4-inch-diameter microphones (M1 and M2), and Brüel & Kjær measuring module "Acoustic Material Testing" are used to measure the absorption of the metasurface [34]. Since the rear of the wall is a hard wall condition, we can assume that there is no transmission. By placing a building block inside the tube, the absorption spectra can be measured for the corresponding metasurface. A digital signal (white noise) powered by the amplifier is sent to the loudspeaker. The absorption coefficient was obtained by analyzing the signal by two microphones. Figure 2(d) shows the comparison of the absorption coefficients obtained by the theoretical analysis, numerical simulation and experiments. One clearly observes that the absorption coefficients obtained by the theoretical analysis and numerical simulation present an excellent agreement with the experimental measurement. It can also be observed that the measured bandwidth is slightly larger than simulated one. This small mismatch arises from the fabrication inaccuracy.

It is worth mentioning that the underlying physical mechanism behind our reported concept is not based on the classical quarter-wavelength resonator theory. The physics behind our absorbing metasurface is based on a hybrid resonance mechanism of coiling-up space geometry due to the coiled chamber (spiral) and labyrinthine passages, and the resonance of Helmholtz-like cavities formed by the corners of the spiral. The total path for the propagating waves inside the coiled chamber is about $\lambda/4.7$, which is significantly smaller than the path allowed by classical quarter-wavelength resonators.

We also would like to emphasize here, that the physical mechanism and the design of the reported absorbing metasurface are significantly different from the reported ones in our previous works [12, 24]. Indeed, although the current design has few similarities with the one reported in our previous researches in terms of using coiling-up space geometry and embedded aperture, the current concept is based on new physics and different architecture than the former, enabling the achievement of unprecedented absorbing properties.

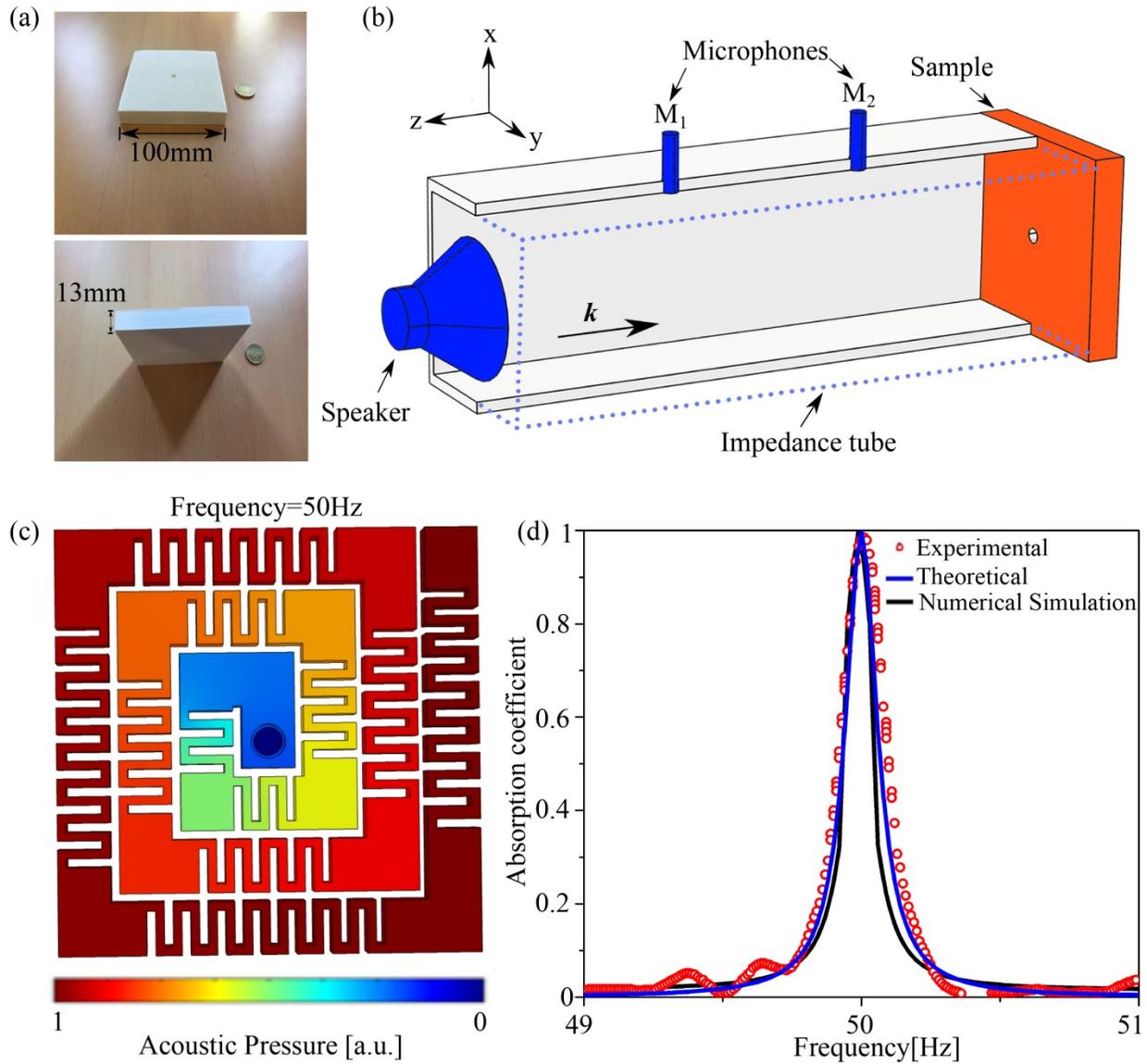

**Figure 2**: (a) Photographs of experimental sample (b) Schematic of the experimental set-up for measuring absorption using the two microphones method. *k* denotes the incident wave vector (c) Simulated sound pressure profile at 50Hz (d) The absorption coefficients of the presented metasurface (shown in the inset photo) with geometrical parameters: $a$=100mm, $d$=6.5mm, $h_a$=9mm, $h_c$=12mm, $w$=12mm, $t$=$g$=2mm. The solid black line, blue line and red dots represent the numerical simulation, theoretical and experimental results, respectively.

As the bandwidth is a highly desired feature when it comes to the absorption in general, and especially in low-frequency regime, we, in follow, introduce an approach by which we can provide a broadband absorbing metasurface operating around a frequency of 50 Hz. For this aim, we design a supercell consisting 9 unit cells (3×3) resonating at different frequencies and having obviously different size aperture diameters as shown in Fig. 3(a) The total thickness of the supercell is still 13mm and the side length of the square supercell is 30cm. Figure 3(b) shows the equivalent electrical circuit for the super cell. As the unit cell are arranged in parallel, total impedance of this super cell is given by,

$$Z_{total}^{-1} = Z_1^{-1} + Z_2^{-2} + \cdots + Z_9^{-1} \qquad (8)$$

The equivalent acoustic impedance $Z_{total}$ at the surface of the sample must be equal to the impedance of the air to achieve full absorption ($Z_{total}=Z_0=\rho_0 c_0$). Figure 3(c) shows comparison of the absorption coefficient of the broadband metasurface obtained by theoretical calculations and numerical simulations. Here, the average absorption is higher than 90% for the given frequency range, but doesn't reach the perfect absorption. This is due to the fact that the hybrid structure still has a small impedance mismatching, making the effective acoustic impedance at the surface slightly deviates from $\rho_0 c_0$. The bandwidth can be further improved by adding more unit cells. The relation between the bandwidth and the number of the unit cells is given in Fig. 3(c), showing almost a linear dependence.

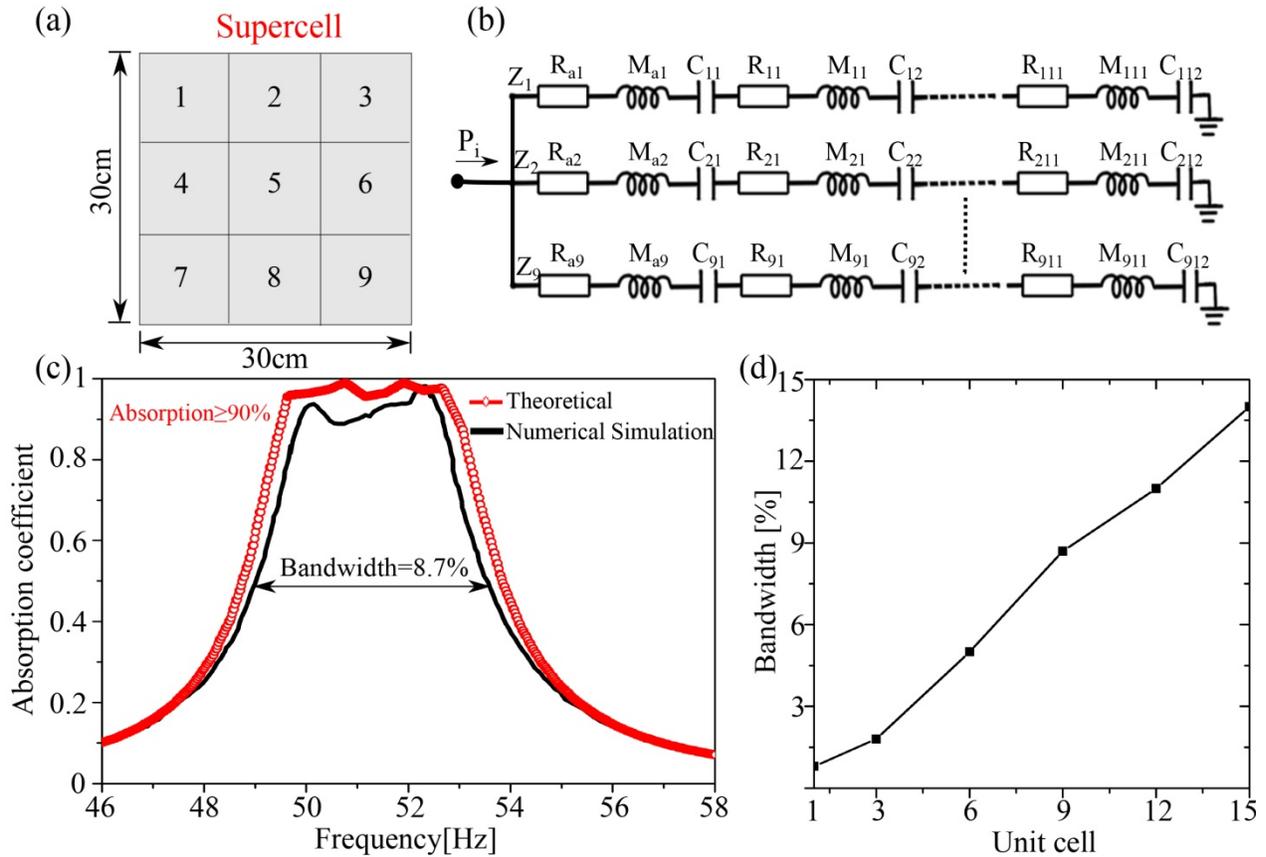

**Figure 3**: (a) The supercell consists of 3×3 unit cells denoted as 1-9. The side length of the supercell is 30cm. (b) Equivalent circuit of the supercell. (c) The absorption for the frequency range of 45-56Hz. 8.7% bandwidth is achieved while preserving the absorption higher than 90% for 3×3 unit cells. The solid black line and red dots represent the numerical simulation and theoretical results, respectively. (d) Relationship between bandwidth and the number of unit cell in the supercell.

We have introduced and demonstrated an unprecedented concept of extreme low-frequency acoustic absorption based on ultra-thin multi-coiled metasurface, and breaking the physics limitation imposed by the quarter wavelength resonator. We analytically, numerically and experimentally have demonstrated the effectiveness of this new physical mechanism and its real added value compared to the previous works. The presented multi-coiled metasurface is capable of fully absorbing acoustic energy in the low frequency regime with a deep sub-wavelength thickness. The full absorption is achieved at 50Hz with the thinnest metasurface achieved until now, 13 mm ($\lambda/527$), overcoming the major obstacles related to intrinsically physical dissipation mechanism and to the huge size of classical absorbers when it comes to absorbing very large wavelengths. To achieve a superior capability of acoustic absorption and broadband feature, we have introduced a new physical mechanism and used a super-cell approach. This proposed metasurface concept introduces a real leap towards pragmatic applications in extreme low-frequency acoustic absorption, and present a genuine breakthrough to develop ultra-thin, lightweight and efficient meta-absorbers.

**Acknowledgments:**

This work is supported by the Air Force Office of Scientific Research under award number FA9550-18-1-7021. The authors also acknowledge the support from la Région Grand Est.